\documentclass{svproc}
\usepackage{graphicx}
\usepackage{footmisc}
\usepackage{amsmath}
\usepackage [autostyle]{csquotes}

\usepackage{url}

\begin{document}
\mainmatter              
\title{\large Towards Eusociality Using an Inverse Agent Based Model }
\titlerunning{Towards Eusociality}  
%
\author{John C. Stevenson }
\authorrunning{JC Stevenson} 
%
%
\institute{Long Beach Institute, Long Beach, NY 11561\\
\email{jcs@alumni.caltech.edu}
}

\maketitle              

\begin{abstract}
	The  emergence of eusocial species is both very rare in evolutionary history and results in remarkably successful species. By inverting an agent based model, agent rules are discovered that display behaviors characteristic of eusocial species as well as other behaviors that lead to unexpected population dynamics. By holding the agents' genome constant across the colony and allowing the agents' rules to evolve, the individual behaviors exhibit phenotypic plasticity in response to environmental cues. The phenotypically driven reduction of intrinsic growth rates and the emergence of non-reproducing phenotypes both demonstrate selection pressure at the colony (system) level. The efficiency of an evolved colony is shown to have a strong relationship to the computational capacity of the agents. Various other emergent behaviors, both eusocial and otherwise novel, are identified and discussed. A path forward to more capable eusocial populations and inter-colony evolution is outlined.

	
\keywords{eusocial, inverse ABM, phenotropic plasticity, stochastic gene simulation, genetic programming, iGSS}
\end{abstract}
\section{Introduction}
Eusocial species represent a very small fraction of the total species on earth and yet they rank among the most ecologically dominant land animals by population and biomass \cite{wilsonN}.  The limited number of species that independently evolve eusociality in diverse taxa suggest this occurrence is a phylogenetically rare event and is considered "one of the great mysteries of biology" \cite{howard}.  The definition of eusociality has changed since its first use in 1966 for nesting bees \cite{batra,michener}; through Wilson's classification as colonies with overlapping generations, division into reproductive and non-reproductive castes, and cooperative care for the young \cite{wilson1971,wilson,ward}; to an explicit definition that tries to incorporate the many eusocial communities in both arthropods and vertebrates \cite{crespi}. For the purposes of this paper, Wilson's classification is unambiguous. Additional eusocial characteristics often found include nesting, environmental effects on reproduction rates, coexistence of different phenotypes, haplodiploiy or similar reproductive strategies, and other cooperative behaviors such as group foraging and defense \cite{freidman}. For those colonies whose reproductive caste is singly mated queens, all the female members of these colonies have very similar genomes; and the diverse physical and behavioral female phenotypes found within the colony are due to responses to each individual's environment (phenotypic plasticity). 

Agent based models, as used in this research, are inherently social. Agents interact with and affect not only the environment but they also compete and cooperate with each other. Classification of biological, sociological, and ecological models include minimal models for systems and synthetic models of systems \cite{rough}. Synthetic models of systems match the macroscopic results of the model to empirical data \cite{bianchi,heckbert,patel,grimm,wiegand} and provide explanatory rules \cite{grimm,wiegand,epstein,brugnera}. The rules are either manually crafted or automatically generated with evolutionary algorithms \cite{holland,koza} such as those used in inverse Generative Social Science (iGSS) \cite{vu,gunaratne,greig}. The resultant macroscopic characteristics from the completed simulation are compared to an objective function \cite{greig} (exogenous selection) and the rules are updated and the simulation repeated until the stopping criteria are met.

In contrast to these synthetic models, a minimal model of a system does not attempt to calibrate to an empirical objective function. Rather, a population of agents freely evolves within an environment under evolutionary selection. Some models in this category apply selection pressure exogenously, for example Iterated Prisoner's Dilemma contests \cite{axelrod,fogel,lindgren,lindgrenNordahl,miller,skyrms,skyrmsB}. Others apply the selection pressure endogenously within the simulation as a \textcquote{gause}{struggle to survive}, where more fit individuals reproduce and replace the less fit. Well-known examples of endogenous selection in a minimal model of a system are Epstein and Axtel's classic Sugarscape \cite{eps:axl}, Pepper and Smuts's alarm calling and feeding restraint model \cite{pepper}, and a demographic Prisoner's Dilemma study \cite{epsteinIP}. When applying evolutionary optimization methodology to these endogenously optimized minimal models, much of the complex algorithmic machinery used for optimizing candidate populations outside of the simulation is not required. 

Within this group of minimal models of systems that evolve rules, some qualify as complex adaptive systems (CAS) which may optimize either on the level of individuals (CAS2) or as a system (CAS1)\cite{wilsonCAS}. CAS2 systems often evolve into a \textcquote{ostrom}{tragedy of the commons} thus stimulating research on cooperation. Game theory is one productive area for this research \cite{axelrod,fogel,lindgren,lindgrenNordahl,miller,skyrms,skyrmsB}, but these games still optimize at the individual level (CAS2). True system level optimization requires individuals to reduce their own survival and reproductive success for the benefit of their community \cite{howard,pepper,wilsonSuper,wilsonCAS}. This research extends the iGSS approach to a minimal model for a system with endogenous evolution of genetically-programmed agent rules. The emergence of phenotypically driven reductions in intrinsic growth rates and of non-reproductive castes suggests that optimization is occurring at the colony level (CAS1). By developing a language and grammar of agent rules independent of the agents' genomes, competition between colonies of different genomes would drive evolution of the queens' genomes, though that is beyond the scope of this paper.

The principal results of this research are the emergence of phenotypic plasticity within a colony of agents with identical and fixed genotypes \cite{west,dewitt} and resultant colony-level optimizations (CAS1). Phenotypic behaviors reduced the intrinsic growth rate of the colony through competitive exclusion (benefiting both the individual and the colony) and through generation of viable populations of non-reproducing phenotypes (sacrificing individual reproductive success). A number of ancillary eusocial behaviors also emerged including stable coexistence of different phenotypes, cooperative foraging by different phenotypes, overlapping generations, phenotypic driven changes in reproduction rates, haploid reproduction, and phenotypes that only breed and do not forage. The only defining eusocial behaviors not observed were cooperative care of the young, and nesting and its defense \cite{howard,wilson1971,wilson,ward,crespi}. A number of other emergent behaviors are also discussed in detail: for their unexpected novelty, for how the population dynamics reflect and drive the evolving agent rules, and for a deeper understanding of how phenotypic behaviors emerge and evolve. These results include evidence of colony fitness proportional to agent computational and memory capacities, competitive exclusion of various phenotypes of interest, and stable population levels well above the predicted steady state carry capacity.\footnote{Colony fitness is inferred from mean population level (realized carry capacity) and volatility. Relative phenotype fitness is determined by competitive exclusion.}


The paper proceeds by first describing the underlying agent based model and its population dynamics with hard-wired agent rules and a genome that contains the relevant agent characteristics.  The language and grammar of genetically programmed rules for the agents are defined, and hand-crafted programs that replicate the hard-wired rules are presented. Random initial agent programs are then allowed to evolve across the various (constant) genetic and computational capacity parameter. These results are discussed both in terms of eusociality and the novel extension of iGSS methodologies to a minimal model of a system. Future research for evolving true eusocial colonies is outlined.

\section{Models and Methods}
\subsection{Underlying Agent Based Model with Genetic Characteristics}

\begin{figure}
	\begin{center}
		\includegraphics[angle=0,scale=0.28]{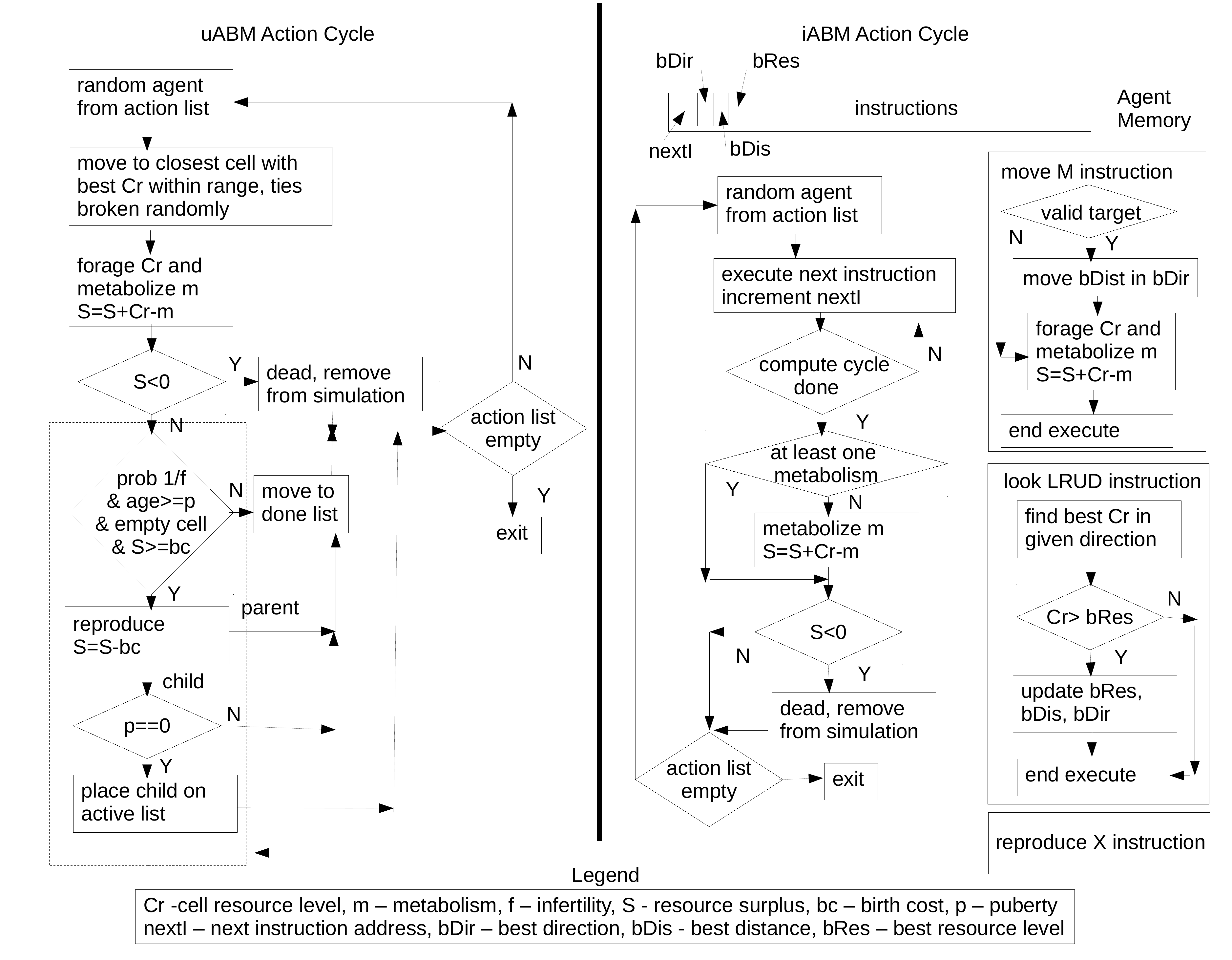}
	\end{center}	
	\caption {Action cycles for the uABM (left) and the iABM (right)}
	\label{fig:action}
\end{figure}

The underlying spatial agent-based model (uABM) is based on a minimum model of a system by Epstein and Axtell \cite{eps:axl}. The agent characteristics that are part of the evolutionary process are defined as genes on a single chromosome which reproduces with occasional mutation (haploid parthenogenesis). These characteristics are stochastic infertility, puberty, birth costs (rather than endowments), and introvert/extrovert preference. The remaining agent characteristics and landscape properties are fixed for each run. The agents interact on an equal opportunity (flat) landscape of resources. The action cycle for the uABM is depicted as a flow chart on the left side of Figure \ref{fig:action}. Detailed descriptions of the ABM parameters and processes sufficient to reproduce the uABM are provided in Appendix A. 

The dynamics that emerge from this simple underlying model have been shown to agree with time delayed logistic growth models for single species \cite{murray,kot,stevenson}, stochastic gene diffusion models \cite{ewens,stevenson}, and modern coexistence theory \cite{barabs,chesson2000,chesson2015,chesson2018,stevensonX}. When an initial population of agents with random heterogeneous alleles is run with mutation and subjected to endogenous selection pressures of survival, the population evolves to one that is dominated by minimum infertility, minimum non-zero puberty, minimum birth cost, and introversion. These alleles represent selection towards the maximum intrinsic growth rate possible (Appendix B). The zero puberty allele is not dominate due to spatial effects of immediate births, and introversion is preferred to avoid local resource competition. The resultant population dynamic is a tragedy of the commons \cite{ostrom}, where the population has almost no resource reserves, mean agent lifetimes are brutally short, and extinctions are common due to environmental degradation, lack of resource reserves, and chaotic population level trajectories \cite{stevensonEcon}. \footnote{Discrete logistic growth equations, for example Eq. 3, generate population level trajectories that range from stable through oscillating and into chaotic regimes based on increasing values of intrinsic growth \cite{murray,kot,may,mayOster,hutch,liz} (see Figure \ref{fig:logG}).}

The uABM provides the structure upon which the genetic programming of the agents' behaviors is implemented. This approach presents a very large solution space of various combinations of infertility, birth cost, introvert/extrovert, and puberty alleles. Based on the cited results with the uABM using hard-wired agent rules and genetically evolving agent characteristics, the genome parameter space is reduced to only infertility and birth cost alleles. Puberty is held constant at one generation and the introvert/extrovert preference is disabled. Computation capacity of the agents adds a third parameter to the space. Haploid reproduction as clones was selected for simplicity (as exemplified by eusocial ant species Mycocepururs Smithii of Hymenoptera:Formicidae \cite{rabeng,himler}).

\subsection{Agent Programming Language and Grammar}
A simple language replicating the uABM agent rules was designed and integrated into an inverse ABM (iABM). Each agent has a 32 character string which contains the registers and instructions which the simulation executes on each agent's action cycle. Five characters are used for registers leaving up to 27 characters for the program.  These instructions are described in Table \ref{table:apl}. The action cycle for this iABM is depicted as a flow chart on the right side of Figure \ref{fig:action}.

\begin{table}[h!]
	\begin{center}
		\begin{tabular}{|c|c|c|c|c|}
			\hline
			Name & Address & Function & Values & Description \\
			\hline
			nextI & 1-2 & register & 05-32 & address of next instruction  \\
			bDir & 3 & register & UDLRZ &  best seen direction (Z$=$no data)\\
			bDis & 4 & register & 0-9 & best seen distance \\
			bRes & 5 & register & 0-9 & best seen resources \\
			inst & 6-32 & program & UDLRMX & executeable instruction\\
			\hline
		\end{tabular}
		\bigbreak
\begin{tabular}{|c|c|c|c|}
	\hline
	Instr & Description & Action/Test & Result\\
	\hline
	U & look up & find cell max resource above $>$ bRes & store in bDir,bDis,bRes\\
	D & look down & find cell max resource below $>$ bRes & store in bDir,bDis,bRes\\
	L & look left & find cell max resource left $>$ bRes & store in bDir,bDis,bRes\\
	R & look right & find cell max resource right $>$ bRes & store in bDir,bDis,bRes\\
	M & move & fetch bDis, bDir, if 'Z' random values  & move bDis,bDir \\
	X & reproduce & space, birth costs allow reproduction & place new agent in cell \\
	\hline
\end{tabular}
\caption{Architecture and Instruction Set for Agent Programming Langugage}
\label{table:apl}
\end{center}
\end{table}

The number of instructions that can be executed per each agent's action cycle, called computation capacity, is part of the parameter space that is surveyed. Foraging gains, metabolic costs, and deaths occur during the move instruction. Since multiple moves may occur during one action cycle, each move instruction triggers foraging at the new location and incurs the metabolic cost. Birth decisions and associated costs occur during the reproduction instruction. If the action cycle ends without at least one metabolic resource cost, one is applied. With this genetic programming approach, the hard-wired rules of the uABM can be replicated with a computation capacity of six steps. These programs (\textquote{classic} phenotypes) contain 6 instructions in one action cycle that look in four directions, move, and  reproduce (e.g. UDLRMX and 23 other versions of look ordering). For hard-wired rules, a tie for the best direction and distance is broken randomly. For the genetic programming version (iABM), the first instruction is equally seeded with each of the four directions. Other than for these ties, the order of look instructions (before a move) is not functionally significant. If a move is targeted to a location that is no longer valid (occupied either due to a random move based on no look data collected since the last move, or from outdated look data from a previous action cycle) the agent does not move. The results, with initially seeded classic phenotypes for simulations spanning the genome alleles, has indistinguishable population dynamics and agent metrics from the uABM. These classic phenotypes often emerge as good solutions and, surprisingly, are sometimes competitively excluded. 

\subsection{Methods}
\begin{figure}
	\begin{center}
		\includegraphics[angle=-90,scale=0.6]{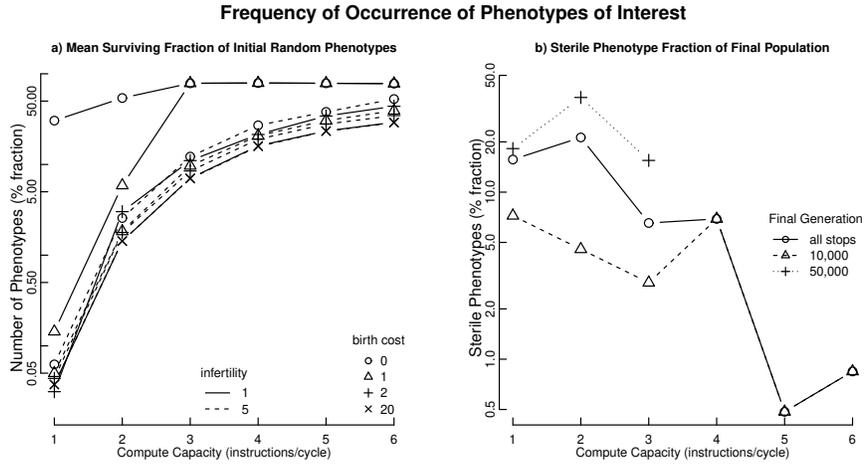}
	\end{center}	
	\caption {Frequency of Occurrence of Phenotypes of Interest} {a) The percent fraction of initial, random phenotypes that survive through the initial population minimum and are fertile across 40 differently seeded runs of 400 initial random phenotypes for each infertility and birth cost by compute capacity. b) The percent fraction of non-reproducing phenotypes at the final generation for all infertility and birth cost alleles by compute capacity and stopping generation.}
	\label{fig:phenoStats}
\end{figure}

All runs are initiated with a population of 400 programs with random instructions of random length. Different seeds generate different initial phenotype populations and resultant population trajectories. Some genetic alleles are so challenging that only a few of the initial random set of phenotypes are able to generate viable, reproducing populations. Figure \ref{fig:phenoStats}a gives the fraction of the initial random population that survives through the initial population minimum and is fertile versus the initial alleles. Each point is a mean fraction of viable surviving phenotypes over 40 differently seeded runs with 400 initial random phenotypes each. Simulation runs are generally stopped at either 10,000 or 50,000 generations, orders of magnitude past the attainment of steady population levels. Phenotype evolution occurs continuously through out the simulations so the stopping point is somewhat arbitrary. Events of interest or long running trends receive longer run times (Figures \ref{fig:phenoStats}b, \ref{fig:sterile} and \ref{fig:length}).

The optimization of the instructions defining the agents' behavior through genetic programming is straightforward. Genetic algorithms \cite{holland,holland2000,booker} and genetic programming \cite{koza} have a large body of techniques for evolving populations to maintain diversity \cite{gupta}, to prevent loss of good solutions \cite{poli}, to improve search efficiency with epigenetic analogues \cite{e-a,lacava}, to improve initial population selection \cite{diaz}, and for multi-objective optimization \cite{rodri}. Since the genetic programs that form phenotypic behavior are selected and propagated continuously throughout the simulations based on a \textcquote{gause}{struggle to survive}, the complex art of exogenous population optimization is avoided. As Figure \ref{fig:phenoStats}a shows, random initial instruction sets over tens of runs are sufficient to generate viable and interesting phenotypes. When an agent reproduces, a single point mutation will occur in the daughter agent at a constant probability $\mu$ per reproduction. If a mutation occurs, a location in the program and a type of mutation are chosen randomly. Three mutation types are implemented: flip to a different random instruction; insert a new random instruction if memory space allows; or knockout the instruction (if the program is longer than one instruction). 

\section{Results}
Broad categories of macroscopic behavior within the parameter space are identified by resultant population dynamics. First, the behaviors that are representative of eusociality are presented and then other emergent behaviors which inform the iABM process are described..

\subsection{Eusocial behaviors}
The emergence of phenotypic plasticity displayed a surprising number of behaviors characteristic of eusocial communities. The consistent emergence of viable, non-reproductive phenotypes is a significant milestone for eusocial behavior. Phenotypically driven changes in growth rate modify the rate set by the colony's genome and result in both higher intrinsic growth rates benefiting the individuals (CAS2) and lower intrinsic growth rates which both benefits the individuals and the colony (CAS1).The ability of different phenotypes to competitively coexist in accordance with modern coexistence theory \cite{barabs,chesson2000,chesson2015,chesson2018} enables most eusocial behaviors. Coexisting phenotypes support caste emergence, cooperative foraging, higher colony fitness, stable sub-populations of sterile phenotypes, and influence population volatility. 

\subsubsection{Populations with Significant Fractions of Non-Reproducing Phenotypes}

\begin{figure}
	\begin{center}
		\includegraphics[angle=-90,scale=0.6]{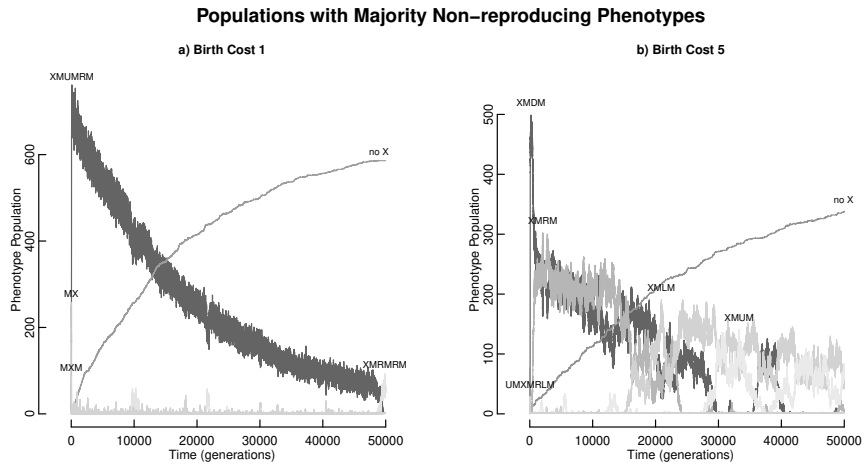}
	\end{center}	
	\caption {Division of Reproductive Labor} {a) The emergence of a large population of non-reproducing phenotypes (no X) for infertility and birth cost 1, mutation rate $0.01$, and computation capacity 2. b) The emergence of a large population of non-reproducing phenotypes (no X) for infertility 1 and birth cost 5, mutation rate $0.01$, and computational capacity 2.}
	\label{fig:sterile}
\end{figure}

Division of reproductive labor is one of the defining characteristics of an eusocial society. Figure \ref{fig:phenoStats}b presents the statistics on the fractions of the final populations which are non-reproducing but viable. These statistics are taken over the constant alleles of infertility and birth cost and presented by compute capacity. Figure \ref{fig:sterile} exemplifies phenotype population trajectories for two representative simulations with differing birth costs where the non-reproducing fraction of the population was greater than half and rose over time. The commonality and viability of these phenotypes suggest two important points. One, there is selective pressure for the emergence of non-reproducing phenotypes which demonstrates system level optimization (CAS1) since these phenotypes never reproduce. These phenotypes consume resources and space without reproducing, which can benefit the colony by reducing the overall intrinsic growth rate, helping to avoid oscillatory and chaotic population trajectories. Though non-reproducing castes are always justified with \textcquote{wilson1971,wilson,ward}{cooperative care of the young}, these experiments suggest that there are other benefits to the colony \cite{howard}. Two, the frequent re-occurrence of this phenotypic behavior as a mutation of a fertile phenotype suggests the fertile phenotypes' rules may carry  neutral pre-adaptation sequences for sterility \cite{miglino}.

\subsubsection{Phenotypically Driven Colony Growth Rates}

Phenotypic behavior can increase or decrease the colony's intrinsic growth rate from that specified by the genome. The uABM replicates discrete logistic growth (with time delay) which has transitions to oscillating and chaotic population level regimes with increasing intrinsic growth rates driven by the allele values. Natural selection at the individual level under these conditions drives toward higher intrinsic growth rates (CAS2). But with a constant colony genome of intrinsic growth, phenotypic behavior often decreases the intrinsic growth rate pushing the colony into more stable regimes\footnote{During the initial growth phase into a rich landscape with few agents, phenotypes are selected for greater intrinsic growth.}. By pushing the colony population dynamics into more stable regimes, the colony benefits by avoiding a tragedy of the commons, chaotic exclusion of more fit phenotypes, and potential extinction (CAS1). Specific examples are discussed in detail in the following section. Adaption of the intrinsic growth rate of a colony to environmental conditions through phenotype plasticity is a characteristic of eusociality.

\subsubsection{Coexistence and Competitive Exclusion}
\begin{figure}
	\begin{center}
		\includegraphics[angle=-90,scale=0.6]{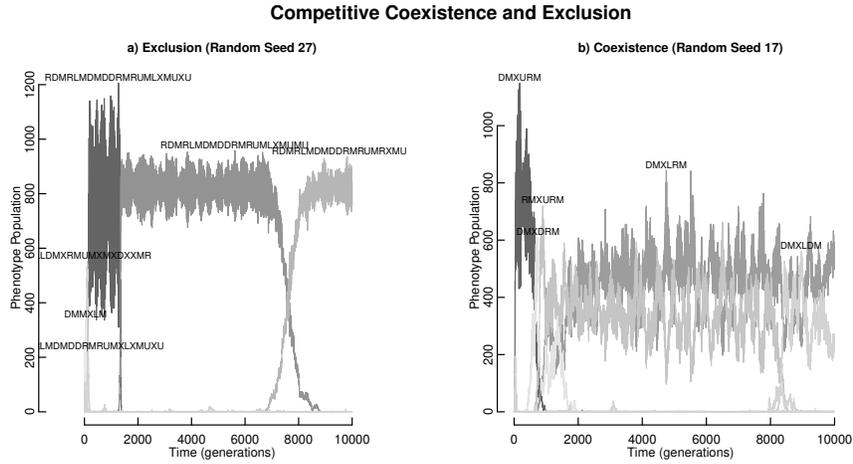}
	\end{center}	
	\caption {Competitive Coexistence and Exclusion} {Two differently seeded solutions that emerge for a constant genome with infertility and birth cost 1, and computational capacity 3.  a) The single mutation of a left look to a right look after generation 7000 drives this invader to exclude the previous resident phenotype. b) Exemplifies coexistence between a phenotype sweeping broadly south with, first, one that looks and moves ENE, and then replaced by a SSW sweeper.}
	\label{fig:Cox}
\end{figure}

In many allele and seed configurations, two or more competing phenotypes will coexist, generating colony fitness that neither would be capable of alone. Other times, a well established resident will be excluded by an invading new mutation. Figure \ref{fig:Cox} presents examples of both. These two examples also provide an excellent demonstration of the wide variety of solutions that will emerge based solely on different seeding of random sets of initial instructions. In both cases the early resident phenotypes have high intrinsic growth rates and generate high population level volatility which, by pushing the dynamics into chaotic regimes, are less fit and are eventually excluded.  The population in Figure \ref{fig:Cox}a with only one resident phenotype has two clear exclusion events where a new mutant invades and quickly excludes a resident population \cite{kang,armstrong}. These new mutants are both single instruction flips. The first exclusion event occurs around generation 1,500 when an reproduction instruction (X) at position 20 mutates to a move 'M' which pushes the population dynamics out of a chaotic regime. The reduction in reproduction with an increase in foraging out competes its parent. The second exclusion event around generation 7,700 is a single mutation at position 16 from a right look (R) to a left look (L). This mutation changed the ratio of left looks to right looks in the phenotype from $\frac{4}{2}$ to $\frac{5}{1}$ producing a relatively more fit left sweeper. In Figure \ref{fig:Cox}b the early resident phenotype is again generating chaotic population dynamics and is again quickly excluded, this time by a pair of phenotypes working together to generate a population that has comparable mean but lower volatility. The first pair of coexisting phenotypes, appearing around generation 2,000, sweeps east-north-east (RMXURM) and broadly south with opportunistic moves to east or west (DMXLRM). At around generation 8,200 a single flip mutation of the broadly south sweeper at position 5 from a right look (R) to a down look (D) proves more fit than the east-north-east sweeper cooperating with the other phenotype. The new pair, parent and child, sweeps both broadly south and south-south-west. Both these paired sweeping patterns are suggestive of cooperative foraging. 

\subsection{Emergent Behaviors that Inform the iABM Process}
Novel population dynamics, not necessarily related to eusociality, provide insight into the iABM evolutionary process. One of the strongest drivers of population dynamics is the computation capacity of the agents. Colony fitness was also shown to be a function of memory size. Classic phenotypes would often emerge and were frequently competitively excluded by other, cooperating phenotypes. Finally a detailed analysis of the emergence of stable populations significantly over the steady-state carry capacity is presented which includes the potential for populations with non-overlapping generations.

\subsubsection{Computational Capacity}
\begin{figure}
	\begin{center}
		\includegraphics[angle=-90,scale=0.6]{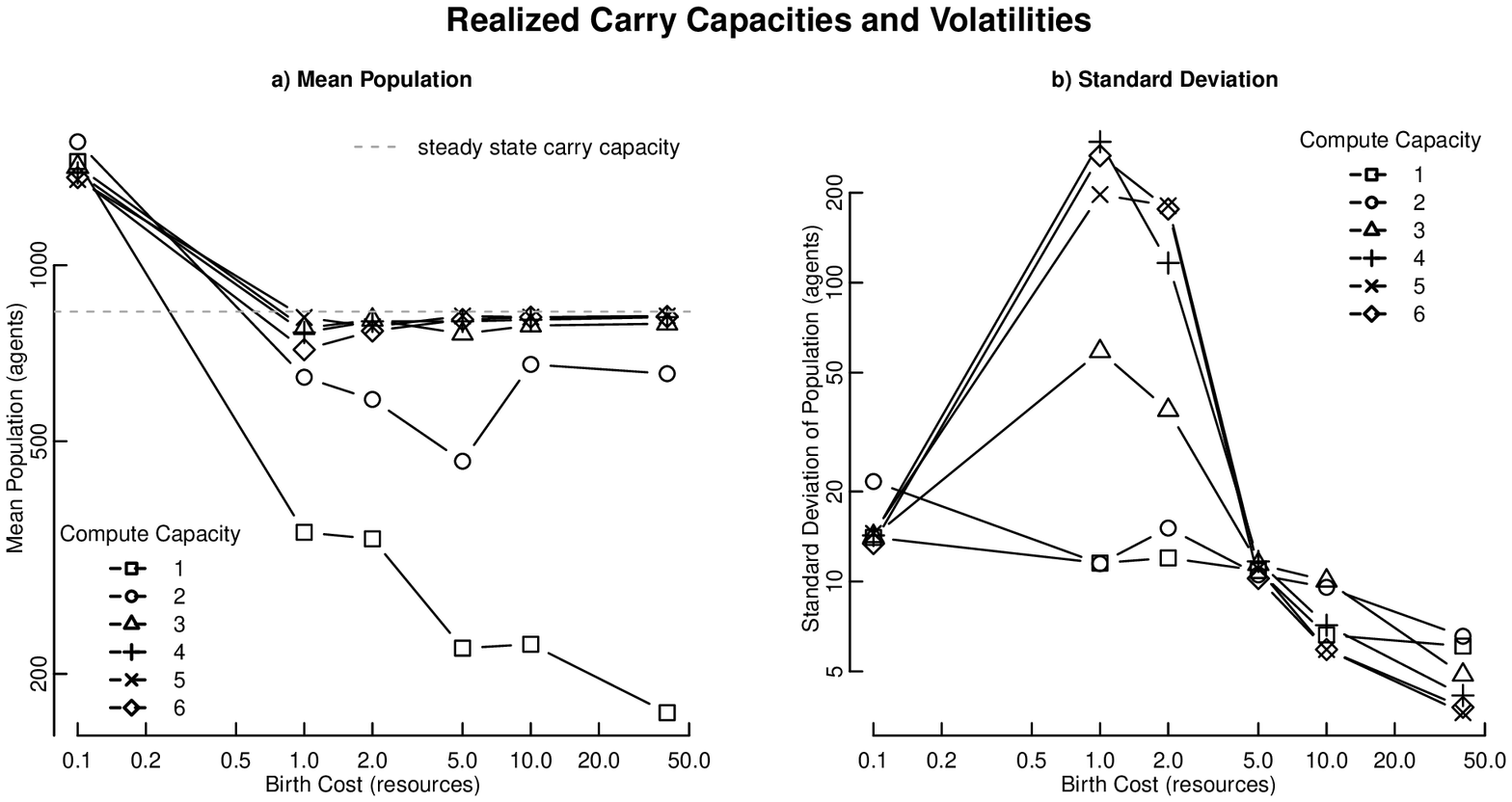}
	\end{center}	
	\caption {Steady State Population Levels with Non-Stochastic Infertility of One} {a) The steady state population levels as proxy for fitness versus birth cost across computational capacities for representative runs. b) The volatility (standard deviation) of these steady state populations levels versus birth cost across computational capacities.}
	\label{fig:cc}
\end{figure}

Emergent agent behaviors were significantly affected by the computational capacity of the agents. Most but not all successful phenotypes have program lengths that are integer multiples of the computational capacity (instructions executed per action cycle), presumably to ensure consistent execution from one action cycle to the next. Surprisingly, stable populations (albeit at varying population levels) emerged for all computational capacities investigated, from one to six. The efficiency (and fitness) of a colony with these phenotypic behaviors has a non-linear relationship with computation capacity as shown in Figure \ref{fig:cc}a. As birth cost increases, colonies with more limited computational capacity become increasingly less efficient. Higher population level variances (Figure \ref{fig:cc}b) indicate that these populations are in oscillatory or chaotic regimes. The surprising stable population levels for birth cost 0 across all the computational capacities are analyzed in Section 3.2.3.

\subsubsection{Memory Size Dependent Fitness}
\begin{figure}
	\begin{center}
		\includegraphics[angle=-90,scale=0.6]{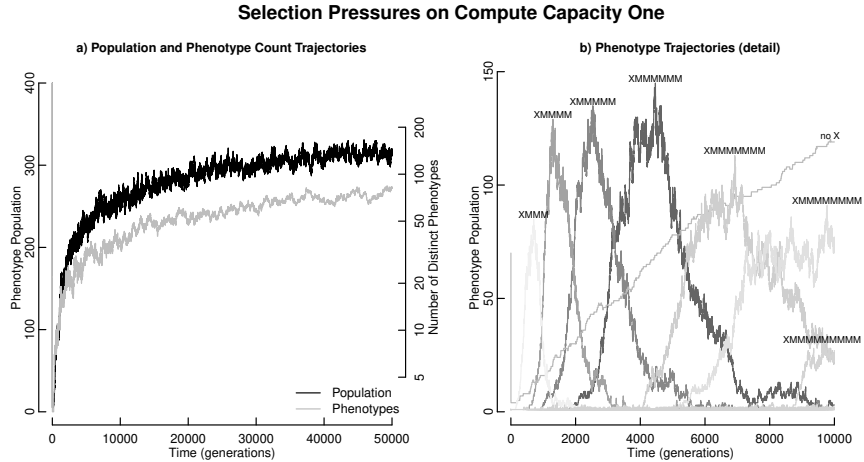}
	\end{center}	
	\caption {Fitness Dependency on Program Length} {a) With infertility 10 and birth cost 5, population levels slowly increasing as mutations drive the program length toward the allocated memory limit. The most fit program is 26 instructions at the end of this run. b) Detail on the phenotype population trajectory highlighting each new, more fit and longer phenotype.}
	\label{fig:length}
\end{figure}

For computation capacity 1 configurations with birth costs greater than zero, a simple strategy dominates the remaining allele spaces of infertility and birth cost. This strategy is of the form 'XMM' with the number of move instructions expanding to eventually fill the agent's allocated program memory. Figure \ref{fig:length}a shows the increase in population level over time as an indication of increasing colony fitness. Figure \ref{fig:length}b shows that increasing fitness is driven by invading mutations with additional move instructions after the initial 'X'.  Any 'XMM' program with more move instructions competitively excludes a similar program with less move instructions. Over time, mutations will discover the next longest program until the memory capacity is reached.

\subsubsection{Steady State Carry Capacity}
Stable population levels well above the steady state carry capacity emerged for colony genomes that contained alleles of birth cost 0 and infertility 1 (non-stochastic) as shown in Figure \ref{fig:cc} for representative runs. Assuming one metabolic cost per action cycle, the steady state carry capacity $K_{ss}$ (agents) is based on the number of agents that can survive on the amount of resources that flow into the landscape each time step $\Delta t$:
\begin{equation}
	K_{ss} = \frac{rcg}{m} = \frac{50*50*1}{3} = 833
\end{equation}
where $g$ is the (constant) rate of resources restored to a landscape cell per $\Delta t$, $m$ is the (constant) number of resources consumed by an agent each $\Delta t$, and $r$ and $c$ are the rows and columns respectively of the toroidal landscape in cells. Figure \ref{fig:cc}a shows that all six computational capacities tested establish stable population levels well above $K_{ss}$ for birth cost 0. 

The assumption that the resources flowing into the landscape are all consumed by surviving members of the population does not apply for these particular populations with levels well above $K_{ss}$. For the compute capacity 1 example, once the population has achieved a steady level, 92\% of the population dies and is replaced. The surviving 8\% of the population forages only 5\% of the resources flowing into the landscape with the remainder metabolized by agents who then die, taking these resources out of the landscape. While from an resource perspective this behavior would seem very inefficient, from the evolving phenotype's perspective population levels are quite high, stable, and these phenotypes successfully invade and exclude other strategies.

\begin{figure}
	\begin{center}
		\includegraphics[angle=-90,scale=0.6]{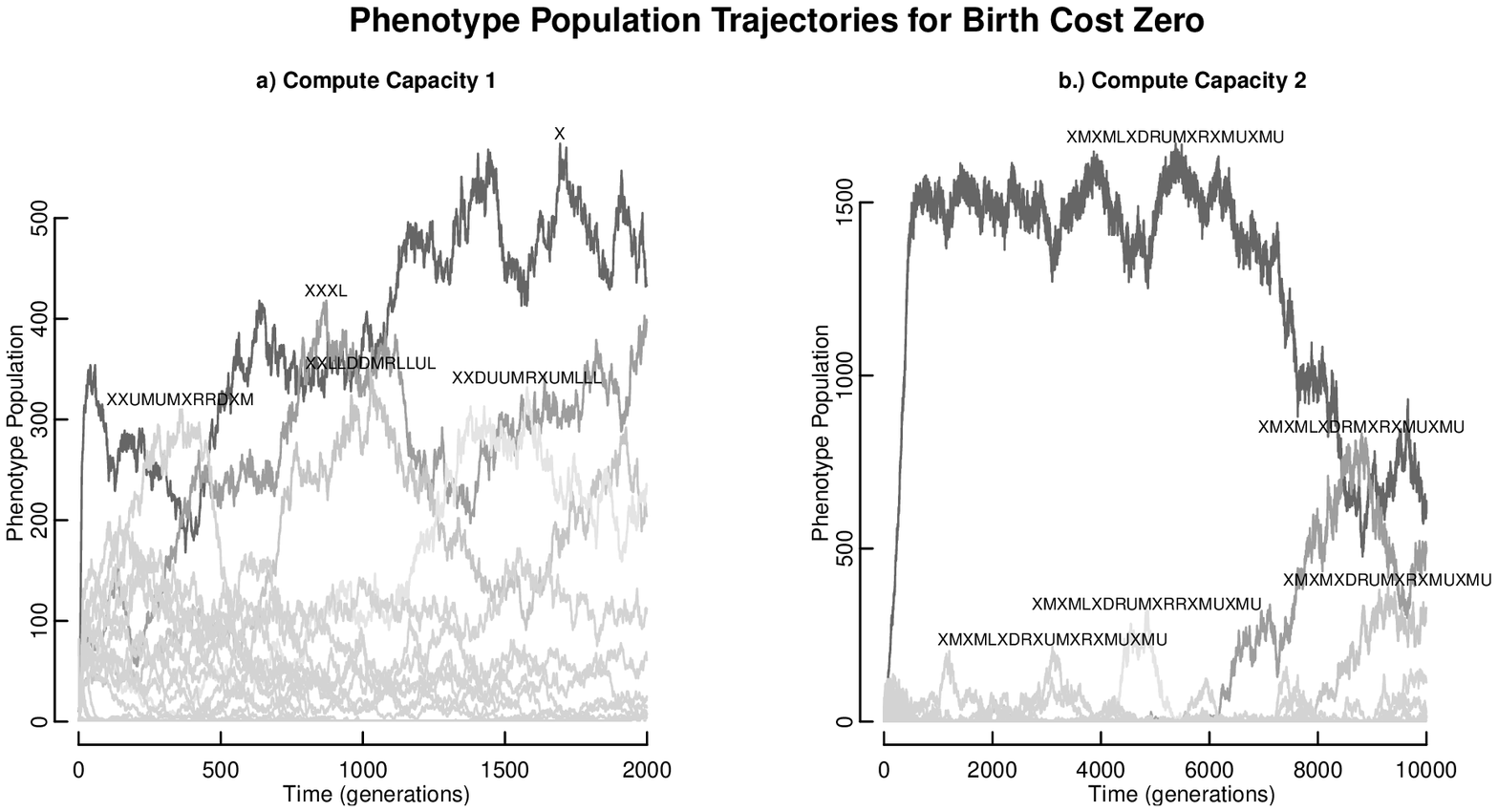}
	\end{center}
	\caption{\small Phenotype Trajectories for Birth Cost Zero} {\small Two representative examples with infertitily 1 and birth cost 0. a) A successful phenotype of the single instruction X emerges with computational capacity 1. b) Foraging strategies emerge  with computational capacity 2.}
	\label{fig:bc0}
\end{figure}

Figure \ref{fig:bc0}a provides the results of a sample run with an initial random program population of 400 agents with infertility 1 and birth cost 0, a probabilistic mutation rate ($mu$) of $0.001$ per reproduction, and with computation capacity 1. The emergence of a single reproduce instruction program was quite surprising since it cannot move to greener fields to forage and the regrowth rate cannot support its metabolism. These viable, single-instruction phenotypes only emerged from the differently seeded initial random population of 400 programs 4 out of 40 times. The population trajectory in Figure \ref{fig:bc0}a also exhibits stochastic diffusion of phenotypes suggestive of neutral selection pressures. Analysis of these programs show that they all execute X for the first two action cycles, which is their maximum life cycle.

\begin{figure}
	\begin{center}
		\includegraphics[angle=-90,scale=0.6]{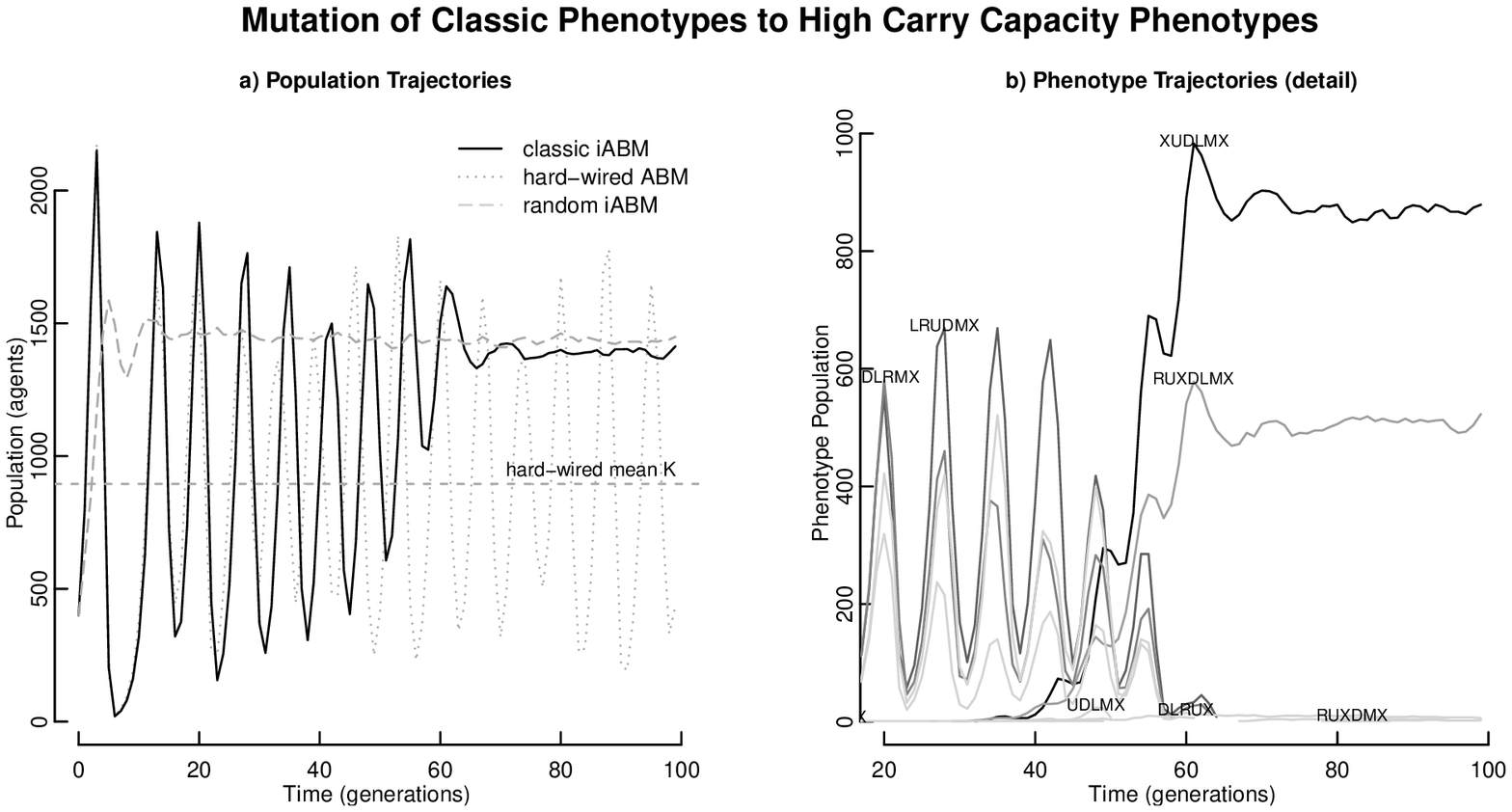}
	\end{center}	
	\caption {Mutation of a Colony Seeded with Classic Phenotypes} {a) The population trajectories for uABM, iABM seeded with classic phenotypes, and randomly seeded iABM; all with infertility 1 and birth cost 0, and computational capacity 6. b) The phenotype trajectories for the classically seeded iABM. The invasion of two mutations with two 'X' per action cycle excludes the classic phenotypes and raises the population level to well over $K_{ss}$.}
	\label{fig:classicBc0}
\end{figure}

For computational capacity greater than 1, successful phenotypes reproduce and forage (moving and perhaps looking) in a single action cycle for faster competitive growth. As long as the initial landscape is rich in resources the agents will carry enough surplus to survive and reproduce and dominate the landscape. Once the population exceeds $K_{ss}$, however, the agents' surpluses will quickly be exhausted and only the phenotypes with a leading 'X' will survive. As an example of this behavior, Figure \ref{fig:bc0}b with a computational capacity of 2 shows the first forage and move phenotype grabs all the surplus, then as it dissipates other functionally equivalent mutants invade and neutral stochastic diffusion occurs. This pattern at a steady population level emerged for all the computational cycles tested, providing a surprising and unexpected population level dynamic. Figure \ref{fig:classicBc0} provides insight on how the uABM did not attain this exceptional carry capacity and how selection pressure mutates the classic phenotype into a higher carry capacity strategy that successfully invades the classic. Figure \ref{fig:classicBc0}a shows the population trajectories for the hard-wired uABM, seeded classic phenotypes in an iABM and randomly seeded iABM, all with infertility 1. Note the classic program with these alleles is in a highly oscillatory regime. The classic phenotype under mutation at first follows the uABM population trajectory but selection pressure quickly identifies more fit programs, and the population trajectory shifts to one similar to the randomly seeded iABM. Figure \ref{fig:classicBc0}b illustrates the mutations that lead to the two new phenotypes that competitively exclude the classic phenotypes and raise the population level well above $K_{ss}$.

While zero birth cost might seem somewhat unlikely given these surprising population levels, conservation of energy and agents is maintained at every time step. To reproduce and then die is a behavior found in octopi, squid, salmon, and mayflies. 


\section{Discussion and Future Work}
The discovery of genetically programmed agent behaviors in a spatial agent based minimal model of a system has demonstrated the emergence of creative and novel agent behavior rules, many relevant to eusocial societies. Phenotypic plasticity, for one,  opens the door for both eusociality and inter-colony evolution. The emergence and viability of non-reproducing phenotypes, a necessary but not sufficient defining behavior for eusociality, demonstrates selection pressure at the colony level (CAS1). Other phenotypically-driven reductions in intrinsic growth rates benefit both the individual (CAS2) and the colony (CAS1).

Many behaviors characteristic of eusocial societies are shown to have emerged from random initial populations of programs whose agents all posses the same colony genome. Cooperation through coexistence leads to higher colony fitness. Non-reproducing phenotypes emerged and increased to majority representation in many colonies. Phenotypes that only bred emerged under birth cost 0 and non-stochastic infertility 1 alleles across all computational capacities tested. Phenotypic plasticity significantly changed the intrinsic growth rate of the colony, moving it into or out of oscillatory or chaotic population regimes. These changes in intrinsic growth rate were often achieved by cooperating phenotypes. 

Numerous examples of distinct, novel and informative agent behaviors based on environmental conditions exhibited phenotypic plasticity. Classic phenotypes emerged for computational capacity 6 but were subject to competitively exclusion by other cooperating phenotypes. Colony population levels well over the theoretical carry capacity were analyzed in detail. One significant effect often not considered for iGSS simulations is the impact computational capacity has on successful strategies. As an example, colony fitness was shown to be proportional to program length for computational capacity 1.

Emergence of conventionally defined eusocial colonies using this model will require the addition of local sharing of resources (cooperative care of young) and sensing local neighbors' colony genome (friend/foe) which, when combined with exploitation of the introvert/extrovert gene, may generate nesting behaviors (philopatry). The current structure of this iABM with a separate queen's genome for each colony coupled with phenotypic plasticity through evolving agent rules supports inter-colony competition and evolution of the colonies' genomes.

%
%

\appendix{ \textbf{ Appendix A - Computational Model and Process}}

Table \ref{table:appA} provides the definition of the agents' and landscape's parameters used for this investigation. Vision and movement are along rows and columns only. The two dimensional landscape wraps around the edges (often likened to a torus). Agents are selected for action in random order each cycle (Figure \ref{fig:action}). The selected agent moves to the closest visible cell with the most resources with ties resolved randomly. After movement, the agent forages and consumes (metabolizes) the required resources. At this point, if the agent's resources are depleted, the agent is removed from the landscape. Otherwise an agent of sufficient age (puberty) then considers reproduction, requiring a lucky roll of the fertility die (infertility), and an empty von Neumann neighbor cell, which are only the four neighboring spaces one step away by row or column. The newborn is placed in the empty cell and either remains inactive until the next action cycle or, if puberty is 0, the newborn is placed on the current action cycle list. With this approach for the action cycle, no endowments for the newborn are required whether for new births or at start-up. Once all the agents have cycled through, the landscape replenishes at the growth rate and the cycle ends.

\begin{table}[h!]
	\begin{center}
		\begin{tabular}{|c|c|c|c|c|}
			\hline
			Agent Characteristic & Notation & Value & Units & Purpose \\
			\hline
			vision & $v$ &  6 &  cells & vision of resources on landscape \\
			movement & -- &  6 &  cells per cycle &  movement about landscape \\
			metabolism & $m$ & 3 & resources per cycle &  consumption of resource \\
			birth cost & $bc$ & 0 & resources &  sunk cost for reproduction \\
			infertility & $f$ & 1-85 & 1/probability & likelihood of birth \\
			puberty & $p$ & 1 &  cycles &  age to start reproduction\\
			surplus & $S$ & 0+ & resources &  storage of resource across cycles \\
			mutation & $\mu$ & $>=0$ & probability & mutation rate \\
			introvert/exovert & $ix$ & 0-2 & true/false/NA & avoidance of crowds \\
			\hline
		\end{tabular}
		\bigbreak
		\begin{tabular}{|c|c|c|c|}
			\hline
			Landscape Characteristic & Notation & Value & Units\\
			\hline
			rows & $r$ & 50 & cells \\
			columns & $c$ & 50 & cells\\
			max capacity &$R$ & 4 & resource per cell\\
			growth & $g$ & 1 & resource per cycle per cell \\
			initial & $R_{0}$ & 4 & resource, all cells\\
			\hline
		\end{tabular}
		\caption{Agent and Landscape Parameters of the ABM}
		\label{table:appA}
	\end{center}
\end{table}

\appendix{ \textbf{Appendix B - Single Species Models from Mathematical Biology}}

\begin{figure}
	\begin{center}
		\includegraphics[angle=-90,scale=0.65]{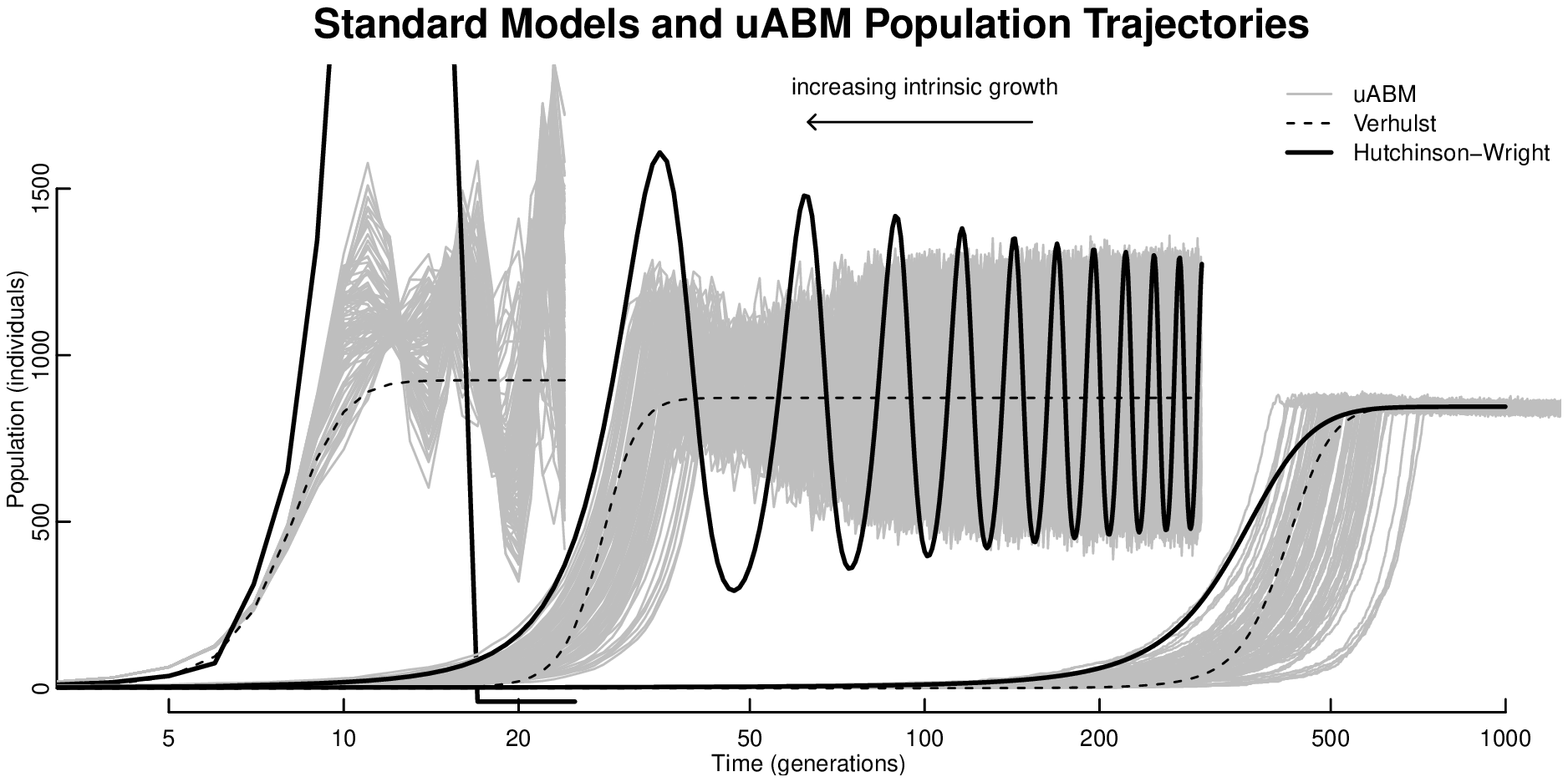}
	\end{center}
	\caption{Population Trajectories} {uABM population trajectories for 100 runs each of infertility 1, 5 and 85 (from left to right), birth cost 0, and puberty 1 with best fit Verhulst (eq. 2) and Hutchinson-Wright (eq. 3) intrinsic growth rates and delay coefficient}
	\label{fig:logG}
	
\end{figure}

A continuous homogeneous model of a single species population $N(t)$ was proposed by Verhulst in 1838 \cite{murray} :

\begin{equation}
	\frac{dN(t)}{dt}=rN(1-\frac{N}{K})
\end{equation}
where $K$ is the steady state carry capacity, $t$ is time, and $r$ is the intrinsic rate of growth. This model represents self-limiting, logistic growth of the population. While the continuous Verhulst Model fits the initial phase of growth well, it does not model oscillating population sizes at the higher rates of intrinsic growth. 
A discrete form of the Verhulst process incorporating an explicit time delay $\tau$ in the self-limiting term was proposed by Hutchinson \cite{hutch} to account for delays seen in animal populations. The resulting discrete-delayed logistic equation \cite{wright}, often referred to as the Hutchinson-Wright equation \cite{kot} is then

\begin{equation}
	N(t+1)=[1+r-\frac{N(t-\tau)}{K}]N(t)
\end{equation}
This model's intrinsic growth rate with $\tau = 5$ captures the steady state, oscillating, and chaotic populations trajectories seen in the uABM with similar intrinsic growth rates. Figure \ref{fig:logG} shows the population trajectories generated by the uABM with specified infertilities; and the continuous Verhulst (eq. 2) and discrete Hutchinson-Wright (eq. 3) trajectories with best-fit intrinsic growth rates and time delay. The regimes of these trajectories move from stable on the right, to steady oscillations in the middle, to fully chaotic on the left based on increasing growth rates.

\end{document}